\documentclass[aps,nofootinbib,prd,eqsecnum,showkeys,reprint]{revtex4-2}
\usepackage{amsmath,graphicx}
\usepackage[colorlinks,linkcolor=blue]{hyperref}

\usepackage{physics,bm,booktabs,caption}

\begin{document}

\captionsetup[figure]{labelfont={bf},labelformat={default},labelsep=space,name={Fig.}}
\captionsetup[table]{labelfont={bf},labelformat={default},labelsep=space,name={Table}}

\title{Realize cosmological inflation in supersymmetric Grand Unified models with $R$-symmetry breaking}

\author{Qian Wan}
\email{wanqian@pku.edu.cn}
\author{Da-Xin Zhang}
\email{dxzhang@pku.edu.cn}
\date{\today}
\affiliation{School of Physics, Peking University, Beijing 100871, China}

\begin{abstract}
	In this paper, we present a discussion for cosmological inflation based on a general renormalizable supersymmetric model which can be naturally embedded into grand unified models. 
	Successful hybrid inflation has been realized with an effective Mexican-hat potential, while avoiding the generation of extra massless multiplets so that gauge coupling unification is maintained in the supersymmetric grand unified models. 
	This is achieved by including $R$-symmetry violating couplings. 
	A relatively large tensor-to-scalar ratio $r\approx 0.03$ and a spectral index $n_s\approx 0.965$ are obtained, which are expected to be further tested by future experiments. 
	Futhermore, The hierarchy of parameters in the model can be solved at the same time, and all dimensionless parameters are in the range from 0.3 to 3 by adjustment of $R$-charges.
\end{abstract}



\maketitle
\tableofcontents
	
\section{Introduction}
The inflationary paradigm \cite{starobinskyNewTypeIsotropic1980,guthInflationaryUniversePossible1981,lindeNEWINFLATIONARYUNIVERSE1982,lythParticlePhysicsModels1999}, as a crucial component of the standard cosmological model, offers a natural resolution to several long-standing cosmological puzzles, including the horizon problem, the flatness problem, and the monopole problem, by assuming that the early universe experienced an epoch of extremely rapid exponential expansion. Since its introduction, the inflationary theory has attracted significant attention. 
In recent years, advancements in observational techniques have propelled the study of inflationary cosmology into a new stage. 
On the one hand, high-precision measurements of the cosmic microwave background (CMB) radiation, exemplified by the Planck \cite{planckcollaborationPlanck2018Results2020} and WMAP \cite{bennettNINEYEARWILKINSONMICROWAVE2013} satellites, have provided robust empirical support for the inflationary scenario. 
On the other hand, the future measurements of CMB $B$-mode, such as the AliCPT \cite{liProbingPrimordialGravitational2019}, CMB-S4 \cite{CMB-S4:2016ple,Abazajian_2022}, and LiteBird \cite{LiteBIRD:2022cnt}, will provide additional data for the study of inflationary theory, thereby enhancing our understanding of the mechanisms and dynamics of cosmological inflation.

Although the inflationary theory has made many important advances in explaining the early evolution of the universe, we still do not understand the mechanism that gives rise to inflation. Therefore, an important and pressing question is how to explain the mechanism of inflation from the perspective of particle physics. 
In other words, how can we naturally introduce the inflaton scalar field within the framework of particle physics. 
Grand Unified Theory (GUT) is the most promising theory to address this issue. On the one hand, the inflaton scalar field can be naturally introduced in GUT. Relevant work can be referred to \cite{araiHiggsInflationMinimal2011,masoudPseudosmoothTribridInflation2020,abidRealisticInflationNoscale2021}. 
On the other hand, cosmological observations indicate that the energy scale of inflation is remarkably close to the GUT energy scale. This suggests that inflation is naturally related to the spontaneous breaking of the grand unified gauge symmetry.

In the attempts to embed inflationary models into GUTs, supersymmetry \cite{ellisCosmologicalInflationCries1982,ellisPrimordialSupersymmetricInflation1983,senoguzTestingSupersymmetricGrand2003,bumseokInflationRealisticSupersymmetric2005,khalilInflationSupersymmetric52011,rehmanSimplifiedSmoothHybrid2015} provides an elegant framework. 
This is primarily due to the numerous advantages that supersymmetry offers in particle physics, such as resolving the fine-tuning problem at the electroweak scale, providing natural dark matter candidates, and facilitating gauge coupling unification. 
Therefore, this paper will focus mainly on the realization of inflation within supersymmetric GUT models. A simple hybrid inflation model \cite{dvaliLargeScaleStructure1994} is
\begin{equation}
	\label{eq1}
	W=\kappa S(\phi^2-M^2),
\end{equation}
where $S$ and $\phi$ are chiral superfields. The superpotential has $R$-charge $2$ so that $R$-symmetry is conserved. The scalar field $S$ (for convenience we use the same symbols for the superfields and the corresponding scalar fields) is the inflaton that drives the inflation, and $\phi$ is the Higgs scalar field that break the GUT symmetry.

However, the realization of inflation in \eqref{eq1} is not natural, as there are three major problems with this model. 
The first problem is that the inflation predictions given by these models are not supported by recent experimental observations, although this can be solved by introducing supergravity corrections and soft supersymmetric breaking terms \cite{shafiObservableGravityWaves2011}. 
The second problem is that the parameter $\kappa\leq 10^{-2}$ is very small and would therefore look unnatural if the above model were embedded in a GUT model without a reason or mechanism to explain the order of magnitude of the parameter in the model. 
The third problem is closely related to $R$-symmetry, where on the one hand, due to $R$-symmetry constraints, the coupling $\phi^3$ is not allowed in the renormalizable superpotential and thus cannot explain the mass splitting of the Higgs multiplets in $\phi$. 
On the other hand, the vacuum expectation value of $S$ is zero when the GUT symmetry is breaking, so there are multiplets that remain massless in the limit of exact supersymmetry \cite{khalilInflationSupersymmetric52011}, which in turn may spoil the gauge coupling unification. 
It is worth noting that similar problems also arise in other attempts to realize inflationary scenarios within the framework of supersymmetric grand unified theories.

Due to the presence of the aforementioned issues, the inflationary models within supersymmetric grand unified theories all appear somewhat unnatural. Therefore, the purpose of this paper is to discuss how to naturally realize cosmological inflation within the framework of supersymmetric grand unified models.

In the present work, we will investigate a hybrid inflation model within the framework of supersymmetry. We first introduce the model at the renormalizable level, including all $R$-symmetry breaking interactions. 
The parameters of the model are constrained experimentally, revealing a significant hierarchy in the parameters which need to be given a natural explaination. 
In analogue to the Froggatt-Nielsen \cite{Froggatt:1978nt} picture for small parameters, the magnitudes of $R$-symmetry breaking effects are assumed to be of the order $(M_X/M_P)^{|R-2|}$, where $R$ is the $R$-charge of a superpotential interaction, and $M_X$ and $M_P$ are the $R$-symmetry breaking scale and the reduced Planck scale, respectively. 
The hierarchy problem in the superpotantial parameters will be naturally explained under this assumption.
 
The structure of this paper is as follows: In Section \ref{section2}, an introduction to our model is provided, and a discussion of GUT embedding is presented, with $SU (5)$ and $SO(10)$ used as examples.
In Section \ref{section3}, we will explore how to realize successful inflation. The inflationary observables are also computed under slow-roll approximation and compared with Planck data. 
In Section \ref{section4}, a short discussion of the viable parameter space that is consistent with current experimental constraints will be given, and we will provide a natural explanation for the hierarchy problem of dimensionless parameters in the model by adjusting the $R$-charges of the superfields. 
Finally, in Section \ref{section5}, a conclusion is drawn.

\section{The Model}
\label{section2}
As discussed earlier, the introduction of $R$-symmetry poses numerous challenges in constructing inflationary models within the framework of supersymmetric grand unified theories. 
Therefore, the starting point of this paper is to assume that $R$-symmetry is approximately broken at the GUT scale. 
It is worth mentioning that $R$-symmetry breaking inflation models have obtained considerable attention. Similar works can be found in \cite{civilettiSymmetryBreakingSupersymmetric2013,khalil1InspiredInflation2019,wanModifiedHybridInflation2025}. 
Consequently, the superpotential is permitted to contain a mildly $R$-symmetry violating coupling $\phi^3$. 
This not only addresses the mass splitting among the multiplets of $\phi$, but as we will see in subsequent discussions, also provides  an elegant solution for other issues.

With $R$-symmetry breaking interactions included, the general form of the superpotential at the renormalizable level can be written as
\begin{equation}
	\label{eq2}
	W=\kappa S(\phi^2-M^2)-\frac{\hat{\mu}}{2}S^2+\frac{\rho}{3} S^3-\frac{\hat{m}}{2}\phi^2+\frac{\lambda}{3}\phi^3,
\end{equation}
where the linear term of $S$ can be cancelled out by an appropriate shift of $S$, thus we will ignore this term in the following discussion. $\kappa$, $\rho$ and $\lambda$ are dimensionless parameters, $\hat{m}$ and $\hat{\mu}$ are parameters with mass dimension but they can be expressed in terms of the dimensionless parameters $m$ and $\mu$ as
\begin{equation}
	\hat{m}=mM_*,\quad\hat{\mu}=\mu M_*,
\end{equation}
where $M_*$ is a typical mass scale of the theory. Considering that we expect to embed the inflationary scenario into the GUT models and $R$-symmetry is approximately broken at the GUT scale $M_G\approx 2.4\times 10^{16}\,\mathrm{GeV}$, it is natural to take all mass scale in the superpotential to be GUT scale, i.e. $M_*=M_G$. For the sake of convenience, in the subsequent discussions we will take the dimensionless parameters to be real and work in Planck units, i.e. the reduced Planck mass $M_P\approx 2.4\times 10^{18}\,\mathrm{GeV}$ is taken to be unity.

The aforementioned superpotential can be naturally embedded into many GUT models, such as $SU(5)$ or flipped $SU(5)$, where $\phi$ is the adjoint representation 24 or 75 of $SU(5)$. 
When embedded into the $SO(10)$ model, $\phi$ can be taken as the adjoint representation 45. It is worth noting that if $S$ is chosen to be 54, the superpotential could generates the DW vacuum \cite{srednickiSupersymmetricGrandUnified1982,babuNaturalGaugeHierarchy1994,babuSupersymmetricSO10Simplified1995,wanExtendedStudySupersymmetric2023}, with the coupling $\phi^3$ being forbidden by symmetry. 
Therefore, our model is expected to provide a natural solution to the Doublet-Triplet Splitting problem in grand unified models. Since this goes beyond the scope of this paper, we will not elaborate on it here. 
Additionally, $\phi$ can be taken as the spinor representation 16 in $SO(10)$, but in this case, $\phi^2$ and $\phi^3$ need to be replaced by $\bar{\phi}\phi$ and $(\bar{\phi}\phi)^2$, respectively.

\begin{figure*}[ht]
	\centering
	\includegraphics[scale=1.6]{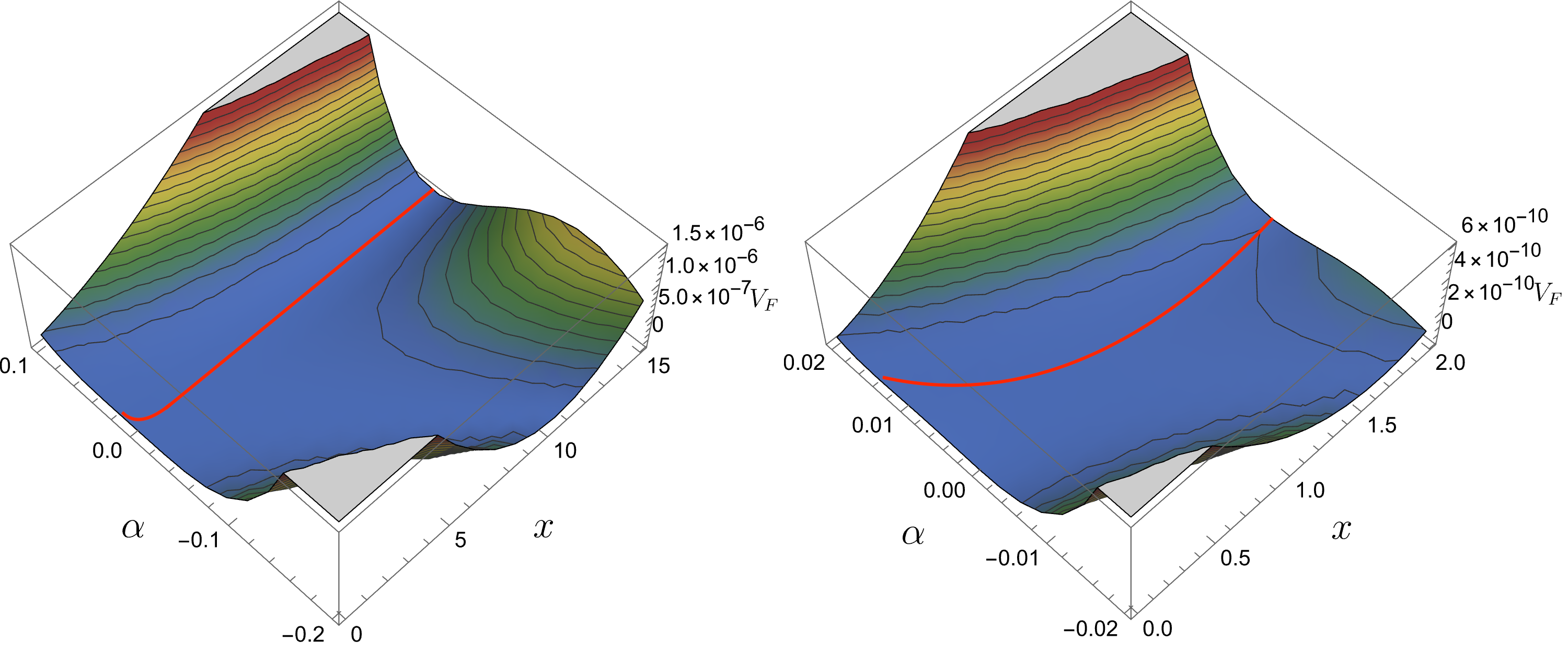}
	\caption{The scalar potential \eqref{eq3} $V_F$ in the $x$ and $\alpha$ plane for the large $x$ region (left panel) and the small $x$ region (right panel), with $\mu=0.474\times 10^{-5}$, $\rho=1.677\times 10^{-7}$, $\kappa=10^{-3}$, $m=10^{-1}$ and $\lambda=10^{-1}$. The red curves represent the inflationary trajectory.}
	\label{Fig1}
\end{figure*}

Working along the $D$-flat direction, the scalar potential $V_F$ is given by
\begin{equation}
	\label{eq3}
	V_F=\left|\kappa\phi^2-\hat{\mu}S+\rho S^2\right|^2+|\phi|^2\left|2\kappa S-\hat{m}+\lambda\phi\right|^2.
\end{equation}
It is evident that the scalar potential described above is semi-positive definite. $\phi=0$ with $S=0$ or $S=\hat{\mu}/\rho$ are supersymmetric vacua. 
However, these vacua do not break the GUT symmetry. Therefore, after the end of inflation, we need to evlove into the following supersymmetric vacua
\begin{equation}
	\label{eq4}
	\begin{aligned}
		\phi_\pm&=\frac{\lambda(\rho\hat{m}-\kappa\hat{\mu})\pm\kappa\sqrt{8\kappa^2\hat{m}\hat{\mu}+\lambda^2\hat{\mu}^2-4\kappa\rho \hat{m}^2}}{4\kappa^3+\lambda^2\rho}, \\ S_\pm&=\frac{\hat{m}-\lambda\phi_\pm}{2\kappa}.
	\end{aligned}
\end{equation}
By performing an appropriate global $U(1)_R$ transformation, the complex scalar field $S$ can be rotated to the real axis, such that $S=x/\sqrt{2}$, where $x$ is the real scalar field driving inflation. 
To discuss the inflationary trajectory, the complex scalar field $\phi$ is parameterized by its real components as $\phi=(\alpha+\mathrm{i}\beta)/\sqrt{2}$. 
In Fig. \ref{Fig1}, we present the explicit form of the scalar potential $V_F$ as a function of $x$ and $\alpha$. Clearly, for a constant value of $x$, $\alpha=\beta=0$ corresponds to local extrema of the scalar potential. 
Thus, when the inflaton field $x$ slowly rolls along the trajectory defined by $\alpha=\beta=0$, inflation occurs. The evolution trajectory of the inflaton field is depicted by a red curve in Fig. \ref{Fig1}.

\section{Inflationary scenario}
\label{section3}
\subsection{General discussions}
According to previous discussions, $\phi=0$ is the inflation trajectory, then the effective inflationary potential will be given by
\begin{equation}
	\label{eq5}
	V_\text{inf}=\frac{1}{4}\rho^2x^2(x-d)^2.
\end{equation}
The parameter $d=\sqrt{2}\hat{\mu}/\rho$ describes the distance between the two global minima along the direction $\phi=0$. 
Clearly, the scalar potential \eqref{eq5} of the inflaton field resembles the Higgs potential that breaks the electroweak symmetry, exhibiting a “Mexican-hat” shape. 
The inflaton field will slowly roll from the unstable supersymmetry-breaking extremum towards the stable supersymmetric vacuum, thereby driving inflation. 
In \cite{croonWessZuminoInflation2013}, a similar scalar potential was obtained using a minimal Wess-Zumino model with a single superfield. 
However, the model discussed in our model employs a GUT breaking trigger field to realize hybrid inflation, which can be directly embedded into GUT models.

\begin{figure*}[ht]
	\centering
	\includegraphics[scale=0.6]{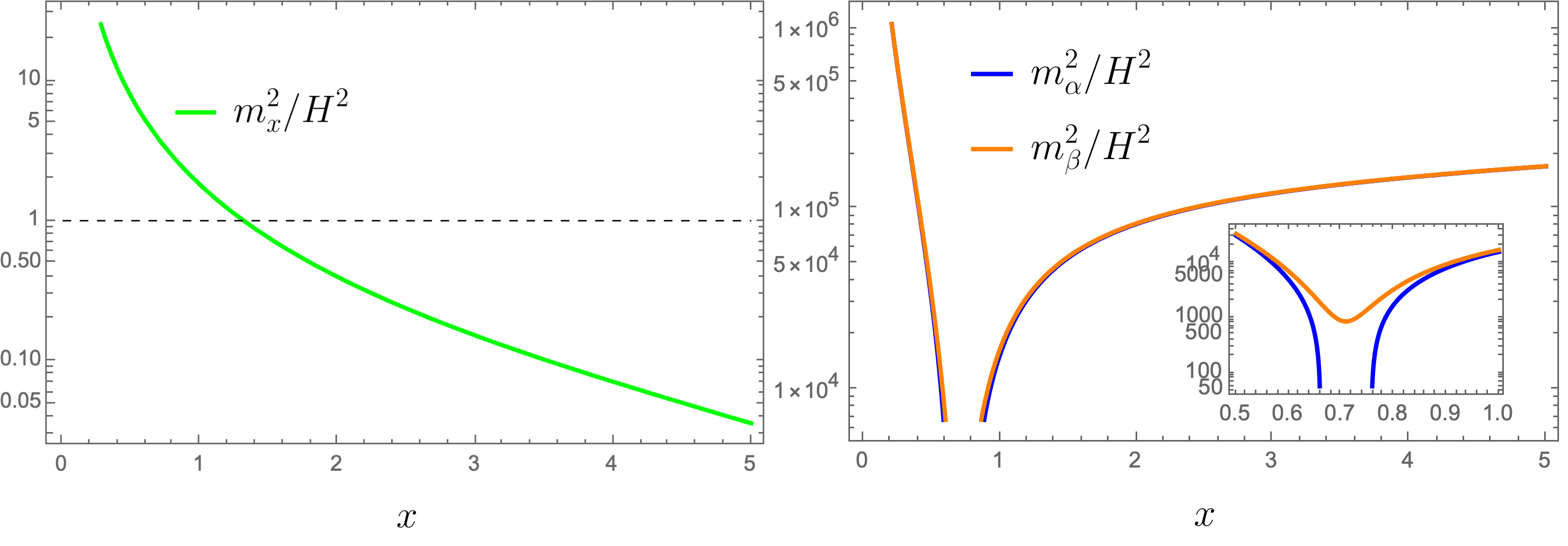}
	\caption{The ratios of the mass squared of the scalar fields $x$, $\alpha$, and $\beta$ to the Hubble parameter squared $H^2$ during inflation, as functions of $x$, with $\mu=0.474\times 10^{-5}$, $\rho=1.677\times 10^{-7}$, $\kappa=10^{-3}$, $m=10^{-1}$ and $\lambda=10^{-1}$.}
	\label{Fig2}
\end{figure*}

We next discuss the stability of the inflationary trajectory. To this end, we need to calculate the masses of the scalar fields during inflation. 
In Figure \ref{Fig2}, we plot the ratios of the squared masses $m_i^2$ of the scalar fields $\alpha$, $\beta$, and $x$ to the squared Hubble parameter $H^2=V_\text{inf}/3$ as functions of $x$. 
It can be seen that during inflation, the scalar fields $\alpha$ and $\beta$ acquire masses significantly larger than the Hubble parameter $H$, and hence they will be frozen at the origin. 
On the other hand, the mass of the scalar field $x$ during inflation is smaller than the Hubble parameter $H$, thus satisfying the slow-roll conditions. 
Additionally, as seen from Fig. \ref{Fig2}, as the inflaton field $x$ rolls towards the origin, the mass squared of $\alpha$ gradually decreases until it becomes negative. The critical point $x_c$ which trigger the waterfall can be computed from $m_\alpha^2=0$, and is given by
\begin{equation}
	\label{eq6}
	x_c=\frac{2\hat{m}+\hat{\mu}+\sqrt{4\hat{m}\hat{\mu}+\hat{\mu}^2-2\rho\hat{m}^2/\kappa}}{\sqrt{2}(2\kappa+\rho)}.
\end{equation}
Then the trigger field $\alpha$ becomes tachyonic and $\alpha=0$ becomes an unstable local maximum (see Fig.\ref{Fig1}), thus $\alpha$ will quickly roll back to the supersymmetric vacuum. The stabilize field $\beta$, however, will remain fixed at the origin during and after inflation.

During inflation, the vacuum energy density is non-zero, which implies that supersymmetry is broken. Consequently, this induces radiative corrections to the tree-level scalar potential $V_F$. 
In the one-loop approximation, the correction is given by \cite{colemanRadiativeCorrectionsOrigin1973}
\begin{equation}
	V_\text{1-loop}=\frac{1}{64\pi^2}\mathrm{Str}\,M_i^4\ln\frac{M_i^2}{\Lambda^2},
\end{equation}
where $\Lambda$ is a cutoff scale. As shown in Fig \ref{Fig2}, the mass squared of the inflaton is less than the Hubble parmeter squared $m_x^2<H^2$, while the masses squared of the stabilize fields $\alpha$ and $\beta$ are $M_i^2\sim 10^5 H^2$ during inflation. 
Given that $H^2=V_\text{inf}/3\sim 10^{-10}$, the one-loop correction $V_\text{1-loop}\sim M_i^4/(64\pi^2)\sim 10^{-13}$ is negligible compared to the tree-level scalar potential \eqref{eq5}.
Therefore, in the subsequent numerical calculations, we can safely ignore the contribution from radiative corrections.

\subsection{Numerical results}
Now we turn to calculate the inflationary observables of the model under the slow-roll approximation and compare them with experimental data to determine the range of parameters in the model. 
The slow-roll parameters can be defined using the higher-order derivatives of the scalar potential with respect to the inflaton field $x$, where the first two slow-roll parameters is given by
\begin{equation}
	\epsilon=\frac{1}{2}\left(\frac{V'}{V}\right)^2=\frac{2(d-2x)^2}{x^2(d-x)^2},
\end{equation}
\begin{equation}
	\eta=\frac{V''}{V}=\frac{2(d^2-6dx+6x^2)}{x^2(d-x)^2}.
\end{equation}
Clearly, the slow-roll parameters depend only on parameter $d$. Under the slow-roll approximation (i.e., $\epsilon$, $\eta\ll 1$), the scalar spectral index $n_s$, the tensor-to-scalar ratio $r$, and the scalar amplitude $A_s$ can be expressed in terms of the slow-roll parameters as 
\begin{equation}
	n_s=1-6\epsilon+2\eta=1-\frac{8(d^2-3dx+3x^2)}{x^2(d-x)^2},
\end{equation}
\begin{equation}
	r=16\epsilon=\frac{32(d-2x)^2}{x^2(d-x)^2},
\end{equation}
\begin{equation}
	A_s=\frac{V}{24\pi^2\epsilon}=\frac{\rho^2x^4(d-x)^4}{192\pi^2(d-2x)^2}.
\end{equation}
The parameter $\rho$ is only determined by $A_s$, and according to Planck 2018 data \cite{planckcollaborationPlanck2018Results2020}, $A_s$ is normalized to $2.137\times 10^{-9}$ at the pivot scale $k_*=0.05\,\mathrm{Mpc}^{-1}$. Finally, the number of e-folds is given by
\begin{equation}
	N=\int_{x_e}^{x_*}\frac{1}{\sqrt{2\epsilon}}\mathrm{d}x,
\end{equation}
where $x_*$ denotes the value of the inflaton field when the pivot scale $k_*$ exits the Hubble horizon during inflation, and $x_e$) represents the value at the end of inflation. Note that during inflation, $0<x<d/2$, hence $\eta<\epsilon$ always holds, and from $\epsilon=1$, we obtain
\begin{equation}
	\label{eq7}
	x_e=\sqrt{2}+\frac{d}{2}\left(1-\sqrt{1+\frac{8}{d^2}}\right).
\end{equation}
It is important to emphasize that to ensure inflation lasts for a sufficient duration, we require $x_e>x_c$. In the subsequent discussions, we will use this condition to constrain the parameters in the model.

\begin{figure}[ht]
	\centering
	\includegraphics[scale=0.6]{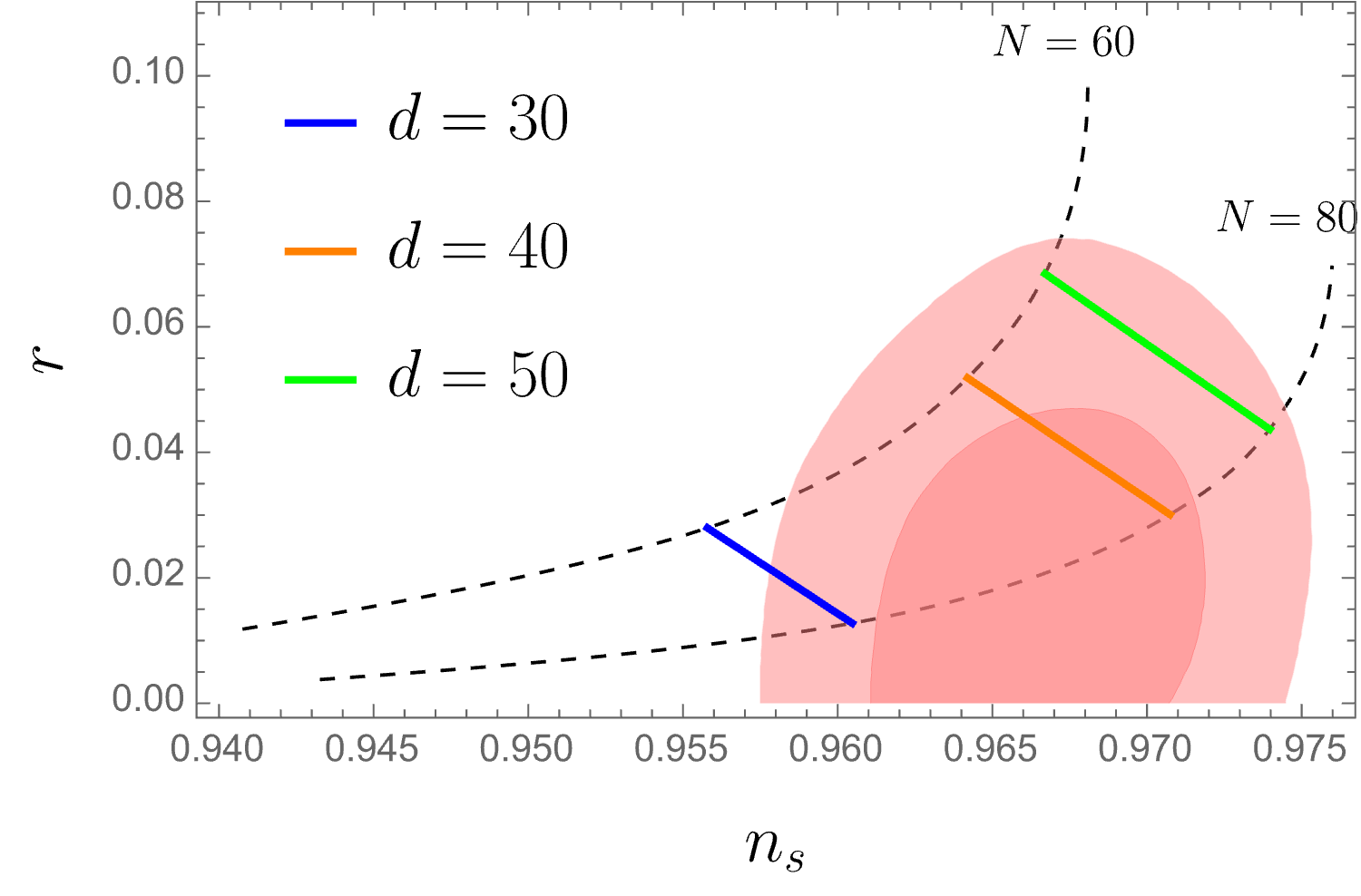}
	\caption{The predictions of the spectral index $n_s$ and the tensor-to-scalar ratio $r$ in the model, compared with the 68\% and 95\% C.L. regions found in analyses of Planck 2018 dataset (Planck TT + low E + BKP + BAO +lensing) \cite{planckcollaborationPlanck2018Results2020}. The black dashed lines represent the model predictions for $d$ in the range from 25 to 90. Additionally, the results for $d=30$ (blue curve), $d=40$ (orange curve), and $d=50$ (green curve), which are consistent with experimental constraints, are also plotted in the figure.}
	\label{Fig3}
\end{figure}

Figure \ref{Fig3} presents the numerical results of the inflationary observables for our model, compared with the constraints from the Planck experiment. 
It can be seen that when $d$ is in the range from 30 to 50, our model yields results consistent with observational data. 
Taking $N=80$ as an example, our model predicts a tensor-to-scalar ratio $r=0.013-0.044$ and a scalar spectral index $n_s=0.961-0.974$ when $d$ is between 30 and 50. 
The results for three typical parameter, $d=30$ (blue), $d=40$ (orange), and $d=50$ (green), are also plotted in the figure. 
It is worth noting that a relatively large tensor-to-scalar ratio $r\sim 0.03$ in our model is highly promising for detection by ongoing and future $B$-mode polarization experiments \cite{liProbingPrimordialGravitational2019,CMB-S4:2016ple,Abazajian_2022,LiteBIRD:2022cnt}, which are expected to achieve a sensitivity of $r\sim 10^{-3}$ at 95\% condidence level.

\begin{figure}[ht]
	\centering
	\includegraphics[scale=0.6]{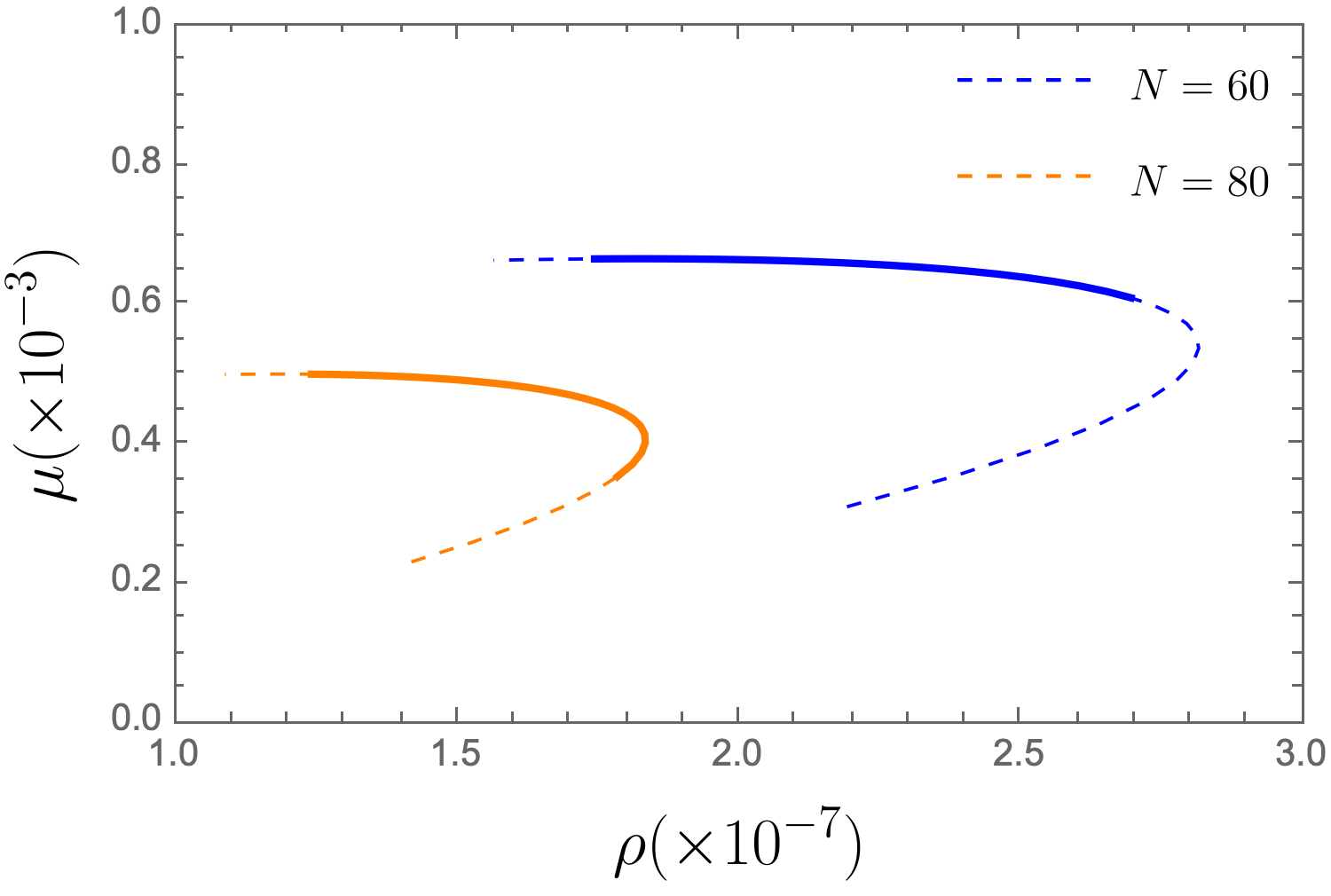}
	\caption{Schematic diagram of the parameter value ranges in the model, where the blue curve and the orange curve correspond to the cases of $N=60$ and $N=80$, respectively, and the solid lines are consistent with the observational data from the Planck experiment.}
	\label{Fig4}
\end{figure}

To facilitate the discussion in the next section, we have plotted the model parameter values that are consistent with experimental observations in Fig. \ref{Fig4}. 
The blue solid line ($N=60$) and the orange solid line ($N=80$) correspond to the Planck experimental data within the 68\% C.L. regions. 
It can be seen from Fig. \ref{Fig4} that the observational results from the Planck measurements impose stringent constraints on the model parameters $\mu$ and $\rho$. 
Specifically, when the e-folding number $N$ is taken to be between 60 and 80, the allowed range of the parameter $\rho$ is $(1.241-2.695) \times 10^{-7}$, and the allowed range of the parameter $\mu$ is $(0.353-0.666) \times 10^{-3}$. 
The reason for such small values of the inflationary model parameters is related to the fact that the inflationary energy scale is much lower than the Planck scale. 
In the next section, we will provide an explanation for this using the breaking of $R$-symmetry.

\section{Hierarchy problem}
\label{section4}
In the previous section, we determined the ranges of the parameters $\mu$ and $\rho$ using the experimental data from the Planck measurements. 
However, the parameters $\kappa$, $m$ and $\lambda$ are related to the predictions of the GUT model and thus cannot be determined by the inflationary observables. 
A detailed discussion of their specific ranges is beyond the scope of this paper. In the following, we will briefly illustrate how to address the hierarchy problem of the model parameters using a simple picture of $R$-symmetry breaking.

Considering that $\phi$ acquires a non-zero vev to break the GUT symmetry, $\phi_+=M_{\text{GUT}}$ serves as a constraint on the parameters. 
Additionally, to ensure that inflation lasts for a sufficient duration, $x_e>x_c$ is required. 
In this paper, we take $\lambda\sim\order{10^{-1}}$, which is a natural choice. Thus a set of dimensionless parameters $\kappa$ and $m$ consistent with this condition have magnitudes of $10^{-3}$ and $10^{-1}$, respectively. 

\begin{figure}[ht]
	\centering
	\includegraphics[scale=0.6]{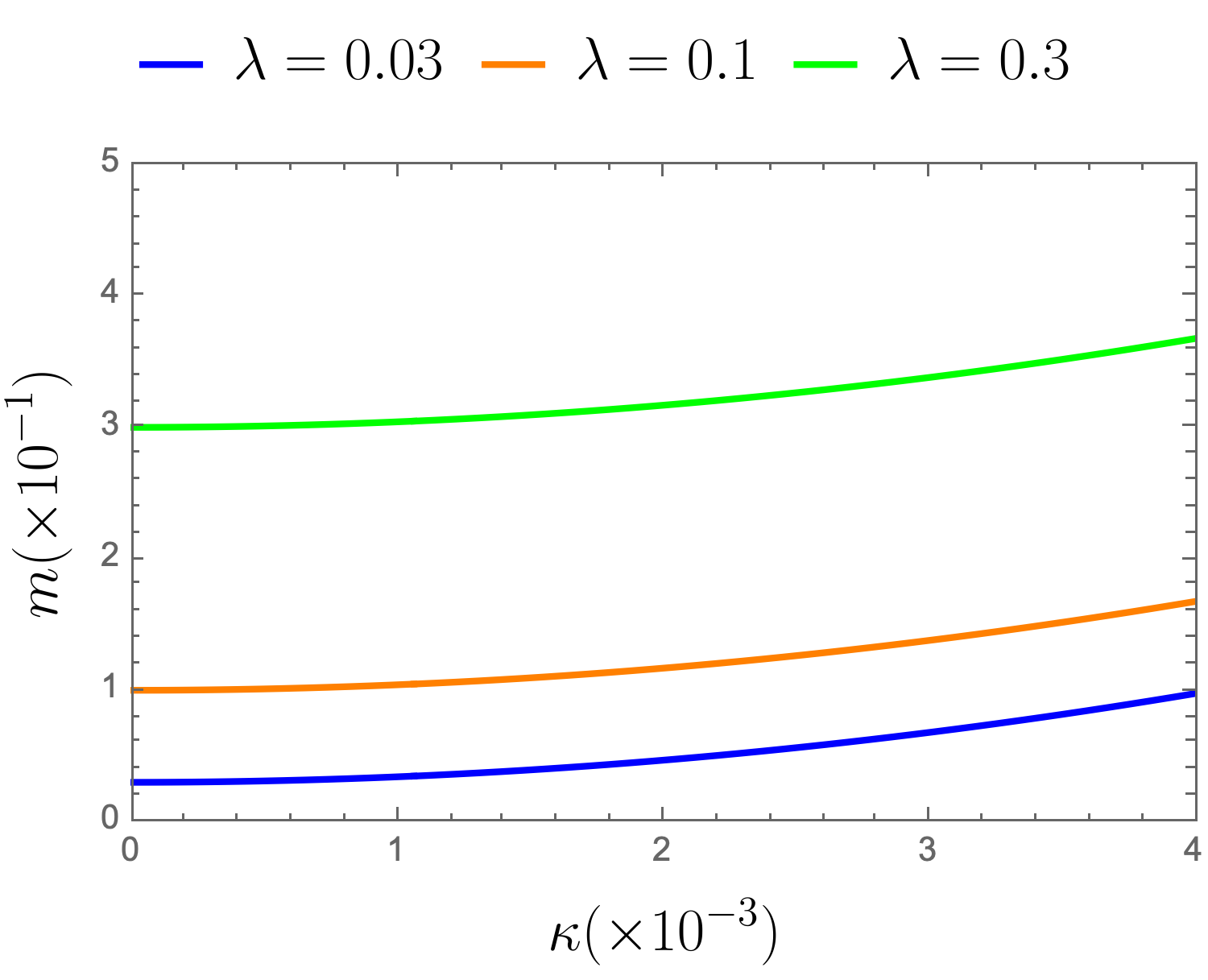}
	\caption{Allowed parameter ranges satisfying the conditions $x_e>x_c$ and $\phi_+=M_G$ when $\lambda\sim\order{10^{-1}}$, where the blue curve, orange curve, and green curve correspond to $\lambda=0.03$, $0.1$, and $0.3$, respectively.}
	\label{Fig5}
\end{figure}

In Fig. \ref{Fig5}, we plot several sets of parameter values that satisfy these conditions, with $\lambda$ taking values of $0.03$ (blue), $0.1$ (orange) and $0.3$ (green). Clearly, when $\kappa$ is in the range of $(0.3-3)\times 10^{-3}$, the allowed range for $m$ is $(0.30-3.38)\times 10^{-1}$.

It can be seen that the parameter ranges in the model vary significantly, spanning from $10^{-7}$ to $10^{-1}$. 
The tiny parameters $\mu$, $\rho$ and $\kappa$ are all related with the inflaton field $S$. 
This hierarchy problem of the model parameters can be explained through the argument on breaking of $R$-symmetry. 

Specifically, we suppose that $R$-symmetry is approximately broken at the GUT scale, allowing for $R$-symmetry violating couplings in the superpotential. Clearly, these $R\neq 2$ operators induce interactions with both $\Delta R=R-2$ and $\Delta R=-(R-2)$
in the lagrangian of equal couplings, hence we assume that the value $|\Delta R|=|R-2|$ measures the magnitude of $R$-symmetry breaking effects,
and a superpotential operator with  $R\neq 2$ is suppressed by a factor $\epsilon_R=(M_X/M_P)^{-|\Delta R|}$, where $M_X$ is the energy scale of $R$-symmetry breaking.
In this sense the superpotential parameters are naturally estimated accordingly, in consistent with the picture of \cite{Froggatt:1978nt,Nir:1993mx}. It is obvious that $R$-symmetry is restored in the limit $R\to 2$. 

Considering the assumption in this paper that $R$-symmetry is broken near the GUT scale, we will take $M_X\sim M_G\sim 10^{-5/2}M_P$. 
In Table \ref{Tab1}, we summarize the model parameters discussed earlier and calculate the suppression factors for different $R$-breaking couplings when $R_S=8/5$ and $R_\phi=4/5$. 
It can be seen that the suppression factors match the orders of magnitude given by experimental parameters, thus providing an explanation for the hierarchy problem in the model parameters.
\begin{table*}[ht]
	\centering
	\caption{The ranges of model parameters consistent with experimental observations and the suppression factors $\epsilon_R$ for $R$-symmetry violating couplings, with $R_S=8/5$ and $R_\phi=4/5$.}
	\label{Tab1}
	\begin{tabular}{ccccc}
		\hline\hline
		Coupling & Dimensionless parameter & Allowed ange & $\Delta R$ & $R$-symmetry violating suppression factor \\
		\hline
		$S\phi^2$ & $\kappa$ & $(0.3-3.0)\times 10^{-3}$ & $6/5$ & $10^{-3}$ \\
		$S^2$ & $\mu$ & $(0.353-0.666)\times 10^{-3}$ & $6/5$ & $10^{-3}$ \\ 
		$S^3$ & $\rho$ & $(1.241-2.695)\times 10^{-7}$ & $14/5$ & $10^{-7}$ \\
		$\phi^2$ & $m$ & $(0.30-3.38)\times 10^{-1}$ & $-2/5$ & $10^{-1}$ \\
		$\phi^3$ & $\lambda$ & $(0.3-3.0)\times 10^{-1}$ & $2/5$ & $10^{-1}$ \\
		 \hline\hline
	\end{tabular}
\end{table*}

Although this paper does not delve into the detailed mechanism of $R$-symmetry breaking, it theoretically provides some clues for studying the breaking of $R$-symmetry and reveals the potential relationship between $R$-symmetry breaking and GUT symmetry breaking. 
After determining the magnitude of model parameters, some predictions for grand unified models can be taken, although these predictions are not elaborated upon here.

\section{Summary}
\label{section5}
In this paper, we discuss a general renormalizable supersymmetric model of hybrid inflation. 
To address the numerous issues arising from $R$-symmetry, we assume that $R$-symmetry is approximately broken at the GUT scale. 
The introduction of $R$-symmetry violating couplings enables the model to successfully realize a slow-roll inflation scenario. 
Our numerical calculations indicate that the model's predictions are consistent with the latest Planck data. 
Moreover, the model predicts a relatively large tensor-to-scalar ratio $r\sim 0.03$ around $n_s\approx 0.965$, which is expected to be further tested by future experiments. 
Additionally, the breaking of $R$-symmetry provides an explanation for the hierarchy of parameter magnitudes in the model. 
By introducing an $R$-symmetry violating suppression factor $\epsilon_R$ and assigning appropriate $R$-charges, the differences in parameter magnitudes can be absorbed into the suppression factor. 
Finally, after GUT symmetry breaking, the model does not produce massless Higgs particles that would disrupt the running of gauge coupling constants. 
The introduced $\phi^3$ term also accounts for the mass splitting among different multiplets of $\phi$. 
In summary, our model can be naturally embedded into supersymmetric GUT models.

\bibliography{document}  

\end{document}